# Dynamic Security Region of Natural Gas Systems in Integrated Electricity-Gas Systems

## Han Gao, Peiyao Zhao, Zhengshuo Li*

*School of Electrical Engineering, Shandong University, 250061, Jinan, China*

*Abstract*—In an integrated electricity-gas system (IEGS), the tight coupling of power and natural gas systems is embodied by frequent changes in gas withdrawal from gas-fired units to provide regulation services for the power system to handle uncertainty, which may in turn endanger the secure operation of the natural gas system and ultimately affect the safety of the whole IEGS. Hence, it is necessary to accurately and efficiently evaluate the dynamic security region (DSR) of the natural gas system in the IEGS by considering the real-time dynamic characteristics of natural gas systems, which are not satisfactorily handled in state-of-the-art works. To bridge this gap, this paper first conceptionally verifies the necessity of the DSR and establishes its mathematical model. Then, a dimensionality reduction method is proposed for the efficient solution and visualization of the high-dimensional DSR evaluation model. A fast evaluation (FE) algorithm is developed to address the difficulties of the nonconvex dynamic constraints in the reduced DSR model. Finally, the necessity and notable advantages of the proposed DSR model and FE are verified based on small and relatively large test systems in comparison with common security region models and algorithms. To the best of our knowledge, this is the first paper that comprehensively presents models and efficient algorithms regarding the DSR of natural gas systems in an IEGS.

## NOMENCLATURE

*Variables*

| | |
|---|---|
| $d_u$ | Gas demand of gas-fired unit $u$ |
| $e_u$ | Output of gas-fired unit $u$ |
| $m_{p,s}$ $m_{p,s}^t$ | Mass flow rate through segmentation points of pipe $p$ |
| $\rho^{(\cdot)}$ | Gas density at nodes or segmentation points of pipes |
| $\gamma_{p,s}^t$ | Lifting variable associated with segmentation points of pipe $p$ |
| $d_{c,t}^C$ | Gas demand of compressor $c$ at time $t$ |
| $d^G$ | Total adjustment amount of gas withdrawal |
| $\Delta d_u^G$ | Adjustment amount of gas withdrawal for gas-fired unit $u$ |

*Sets*

| | |
|---|---|
| $j \in \mathbb{J}$ | Set of nodes in gas network |
| $c \in \mathcal{C}$ | Set of nodes with compressors |
| $g \in \mathcal{G}$ | Set of gas wells |
| $u \in U$ | Set of gas-fired units |
| $p \in \mathbb{P}$ | Set of pipes in gas network |
| $s \in \mathbb{S}_p$ | Set of all segmentation points of pipe $p$ |
| $t \in \mathbb{T}$ | Set of time intervals |

*Parameters*

| | |
|---|---|
| $\alpha$ | Speed of sound (m/s) |
| $D$ | Diameter of natural gas pipe (m) |
| $d_l$ | Demand of natural gas load $l$ |
| $f$ | Friction factor of natural gas pipes |
| $F_u(\bullet)$ | Gas consumption function of gas-fired unit $u$ |
| $\underline{\rho}_j^J, \overline{\rho}_j^J$ | Lower/upper bounds for density at node $j$ |
| $v_g^G$ | Output of natural gas well $g$ |
| $\Gamma_c$ | Ratio of compressors in natural gas network |
| $n_p^{seg}$ | Number of segmentation points in pipe $p$ |
| $k_G$ | Value of lost load in natural gas systems |
| $a_c^C, a_c^k$ | Natural gas consumption coefficients of compressor $c$ |
| $L_{\min}$ | Small range for the system linepack level to vary |
| $\Delta x_p$ | Length of natural gas pipe segments in pipe $p$ |
| $\Delta t$ | Time step |
| $\hat{d}_u^G$ | Dispatch scheme of gas-fired unit gas withdrawal |
| $\beta$ | Participation factor for gas withdrawal changing |

## 1. Introduction

NOWADAYS, there is increasing interaction between power systems and natural gas systems, and the two systems are increasingly collectively considered an integrated electric and gas system (IEGS) [1]. The natural gas demands of gas-fired units play a significant role in the natural gas industry and notably impact natural gas system operations [2][3]. Flexible gas-fired unit output is constantly revised to handle uncertain renewable energy injection in power systems. However, since gas storage is prohibitively expensive [4], most gas-fired plants do not store fuel on site, and natural gas is withdrawn in real time from natural gas pipelines [5],[6], which raises several issues. Specifically, the uncertain gas withdrawal of gas-fired units may result in pressure fluctuations in gas nodes and even violations of natural gas security constraints [7][8]. This in turn may trigger a cascade of events, including fuel shedding in gas-fired units and shortages in the electricity supply [9]. Therefore, it is essential to accurately evaluate the security region (SR) of the natural gas system in IEGSs (i.e., the gas withdrawal range of gas-fired units without violating natural gas security constraints) to ensure the reliable operation of IEGSs.

---
\* Corresponding author.
  E-mail address: zsli@sdu.edu.cn (Z. Li).
  Han Gao and Peiyao Zhao contributed equally to this work and should be considered co-first authors.

However, accurately evaluating the SR of a natural gas system is challenging due to the complex power-gas coupling and natural gas system dynamics. Properly taking into account the unique characteristics of natural gas systems in IEGSs is critical to achieving accurate SR evaluation. To that end, the following issues should be taken into consideration:

**Issue-1 (dynamics introduced by power-gas coordination)** Gas-fired units are widely used to balance the fluctuating injection in power systems, and in turn, this gas consumption of units incurs fluctuating gas withdrawal in natural gas systems [10]. However, unlike steady-state power systems, natural gas systems with slow dynamics typically cannot reach steady-state within intraday operation intervals, e.g., 15 minutes [6]. Therefore, the SR of natural gas systems in IEGSs should be evaluated with consideration of the dynamics of natural gas systems.

**Issue-2 (coupling of gas withdrawal for gas-fired units)** The output of the generating units changes affinely according to the regulation of automatic generation control (AGC)[1] to maintain the real-time power balance [11]. Accordingly, the gas consumptions of the units are interrelated, and the coupled gas withdrawal of gas-fired units should be considered when evaluating the SR.

**Issue-3 (heterogeneous management mechanisms)** Dispatch schemes for gas wells in natural gas systems typically have a time granularity of 1 hour [12], whereas those for generators in power systems have a finer granularity of 15 minutes or less [13]. To accurately capture gas withdrawal changes and be able to guide the security operation of power systems, the SR should be dynamically updated every 15 minutes (i.e., the time granularity of the continuously revised dispatch scheme for power systems).

**Issue-4 (fast evaluation)** Naturally, efficient evaluation algorithms are urgently needed to achieve dynamic updating of the SR in natural gas systems, which is crucial for ensuring online security operations. However, the nonconvex and NP-hard dynamic models of SRs for natural gas systems present significant challenges in developing fast evaluation algorithms [14].

To the best of our knowledge, there is very limited related literature available on SR analysis for natural gas systems in

IEGSs, such as [15]-[20]. Ref. [15] studied the steady-state security region (SSR) of natural gas systems and illustrated the approach of security assessment for natural gas systems based on the SSR. Chen et al. [16] proposed the concept of SSR for IEGSs, and the convex hull-based robust SSR of IEGSs considering uncertain wind power was further presented in [17]. Su et al. [18] first proposed and demonstrated by numerical studies that the SSR for IEGSs is nonconvex and composed of several disjoint components in certain scenarios. Ref. [19],[20] extended the SSR concept to electricity-gas-heat integrated energy systems. However, none of the above works have studied the dynamic updating of the SR, as they all utilize steady-state natural gas system models.

Moreover, the above works mainly focus on how to accurately evaluate the SSR, but most of the above works ignore the requirement for fast evaluation. The SR varies dynamically as the natural gas system states change over time, so it is indispensable to study fast evaluation algorithms for the SR considering natural gas system dynamics.

To bridge the above gaps, this paper proposes the concept of the dynamic security region (DSR) for natural gas systems, establishes its mathematical model, and designs a fast evaluation algorithm. There are two main contributions of our work.

The first contribution aims to solve **Issue-1** and **Issue-2**. This paper proposes the concept and mathematical model of the DSR by considering the natural gas system dynamics and the coupling of gas withdrawal for gas-fired units. To the best of our knowledge, this is the first time that natural gas system dynamics and the coupling of gas-fired unit gas withdrawal have been considered in a related study. Therefore, the phenomena related to natural gas system dynamics that are not captured by the traditional SSR can be reflected by our DSR.

The second contribution aims to solve **Issue-3** and **Issue-4**. Since nonconvex natural gas system dynamic constraints are included, the DSR model is NP-hard. This paper proposes a fast evaluation algorithm (FE) to satisfy the dynamically updated demand of DSR. The rank minimization algorithm with semidefinite programming (SDP) is proposed in the FE to relax the nonconvex DSR model into a tight convex model, and an optimal bisection algorithm is proposed to search for the optimal solution of the convex model.

The remainder of the paper is organized as follows. In Section 2, the concept and definition of DSR are proposed, and a small case is constructed to verify DSR's necessity. In Section 3, the mathematical model of the DSR is proposed. In Section 4, the FE for the dynamic evaluation of DSR is introduced. In Section 5, the validity of our method is verified. Section 6 concludes this paper.

## 2. Concept and necessity for DSR

### A. Definition of DSR

The DSR is defined as a secure operation range for natural gas systems in which natural gas can be withdrawn from the natural gas systems by gas-fired units without violating the dynamic natural gas system security constraints.

---

[1] The automatic generation control is a control scheme that maintains the real-time power balance by distributing the power imbalance to each generator according to the corresponding participation factor [11].

## B. Illustrative example to compare the SSR and DSR

The 3-node natural gas system in Fig. 1 is adopted to demonstrate the impact of natural gas system dynamics on the SR and to verify the necessity of the DSR. As mentioned in Section 3, this paper considers numerous factors in relation to the DSR, among which natural gas system dynamics are of paramount importance. To clearly demonstrate the necessity of dynamics in the SR of natural gas systems in IEGS, only the dynamics are considered in this example. Specifically, the SSR is obtained according to the method in [15], and the DSR is obtained by directly replacing the steady-state constraints in the SSR in [15] with dynamic constraints. Gas-fired units draw natural gas from nodes N1 and N2, and the detailed parameters of the test system can be found in APPENDIX A. The SSR and DSR are evaluated directly by using IPOPT [21].

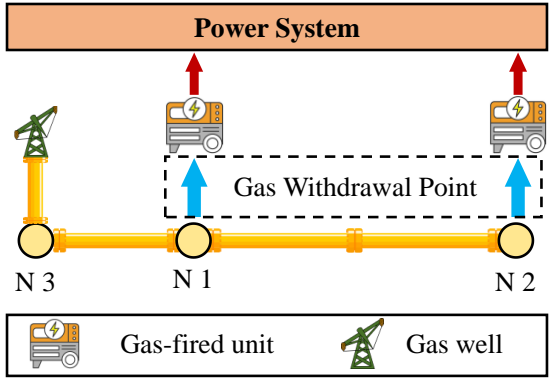

Fig. 1 Schematic of the 3-node natural gas system

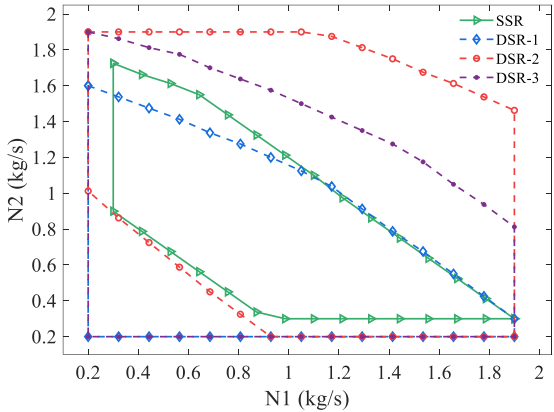

Fig. 2 SSR and DSRs for the 3-node test system. DSR-1, DSR-2, and DSR-3 are DSRs for the test system with different initial states (i.e., gas density and gas flow distribution). The horizontal and vertical coordinates are the normalized values of natural gas withdrawal at nodes N1 and N2, respectively, with a base value of 100.

Fig. 2 reveals the following interesting phenomena. First, DSRs with different initial states are distinct from each other, e.g., DSR-1, DSR-2, and DSR-3 in the figure, and they are not similar polygons as one might have expected, which means that careful consideration of the impact of initial states is necessary. Second, the SSR and the three DSRs are significantly different in terms of shape and area. In Fig. 2, the SSR's area is the smallest, which might be attributed to its inadequate accounting of the linepack[2] effect. However, this smallest area does not mean that the SSR is a conservative or "safe" surrogate of the DSR because certain points within the SSR do not correspond to points in the DSR, potentially resulting in gas withdrawal from the SSR that could violate natural gas system security constraints.

These findings showcase the necessity of considering natural gas system dynamics for the DSR.

## 3. Mathematical models for the DSR

### A. Assumptions

Before introducing the DSR model, three key assumptions for our work are made.

1) **Assumption-1** Traditional gas loads are given and fixed, and only the variation in gas withdrawal of gas-fired units needs to be considered in DSR. The reason for this assumption is as follows.

   Traditional natural gas loads are often predictable, and the users withdraw gas according to the predicted loads, so the uncertainty can be assumed to originate from the gas demand of gas-fired units only [22][24].

2) **Assumption-2** The gas withdrawal of gas-fired units during intraday operation is adjusted based on a day-ahead scheme, and there is a linear relationship between the adjustment amounts of the gas withdrawals, which can be described according to the regulation of AGC. The reason for this assumption is as follows.

   Under the regulation of AGC, the output of the units changes affinely to maintain the real-time power balance [11]. In addition, the relationship between the output and gas consumption of gas-fired units is usually approximated as a linear function [25],[26] so the adjustment amounts of the gas withdrawals of the gas-fired units are linearly related.

3) **Assumption-3** During the natural gas system's intraday operation, the compression ratio of compressors and output of gas wells is known and should not be corrected. The reason for this assumption is as follows.

   Natural gas flow is transported through pipelines at approximately 20 miles/hour, so adjusting the gas wells' output may not be sufficient to handle the load changes [27]. Moreover, it is challenging for the natural gas system operator to optimize complex and dynamic natural gas systems and change the controls in real time [6]. Therefore, in current practice, the dispatch scheme of natural gas systems (e.g., the compression ratio of compressors and output of gas wells) is determined during the day-ahead dispatch rather than redispatching during the intraday operation of natural gas systems [6],[25].

### B. DSR formulation

In this part, we construct a DSR model considering natural gas system dynamics and current industry practices. Considering the time granularities of power systems that are coupled with natural gas systems, the DSR is dynamically updated every 15 minutes. The dynamic natural gas system security constraints that the DSR should satisfy are as follows.

---

[2] Natural gas pipelines can store natural gas internally and can sustain unbalanced operations caused by injection and withdrawal. The gas stored is referred to as linepack [5,16].

$$\text{I} \quad \frac{\rho_{p,s+1}^{t} - \rho_{p,s}^{t}}{\Delta x} + \frac{(m_{p,s}^{t})^2 / \rho_{p,s}^{t}}{\pi^2 D_p^5 / 8f} = 0,$$

$$\forall p \in \mathbb{P}, \ s \in \mathbb{S}_p, t \in \mathbb{T} \tag{1.a}$$

$$\frac{\rho_{p,s}^{t+1} - \rho_{p,s}^{t}}{2\Delta t} + \frac{\rho_{p,s}^{t+1} - \rho_{p,s}^{t}}{2\Delta t} + \frac{m_{p,s+1}^{t+1} - m_{p,s+1}^{t+1}}{(\pi D_p^2 / 4)\Delta x}$$

$$= 0, \forall p \in \mathbb{P}, \ s \in \mathbb{S}_p, t \in \mathbb{T}, t+1 \in \mathbb{T} \tag{1.b}$$

$$\frac{\rho_{p,s+1}^{1} - \rho_{p,s+1}^{0}}{2\Delta t} + \frac{\rho_{p,s}^{1} - \rho_{p,s}^{0}}{2\Delta t} + \frac{m_{p,s+1}^{1} - m_{p,s}^{1}}{(\pi D_p^2 / 4)\Delta x}$$

$$= 0, \forall p \in \mathbb{P}, \ s \in \mathbb{S}_p \tag{1.c}$$

$$\underline{\rho}_j^J \le \rho_{j,t}^J \le \overline{\rho}_j^J, \ \forall j \in \mathbb{J}, \ t \in \mathbb{T} \tag{1.d}$$

$$\rho_{j,t}^J = \rho_{p,1}^t, \ \forall j \in \mathbb{J} \setminus \mathbb{C}, \ p \in \mathbb{P}_j^{fr}, t \in \mathbb{T} \tag{1.e}$$

$$\rho_{j,t}^J = \rho_{p,n_p^{seg}}^t, \ \forall j \in \mathbb{J}, \ p \in \mathbb{P}_j^{to}, t \in \mathbb{T} \tag{1.f}$$

$$\rho_{p,1}^t = \Gamma_j \bullet \rho_{j,t}^J, \ \forall j \in \mathbb{C}, \ p \in \mathbb{P}_j^{fr}, t \in \mathbb{T} \tag{1.g}$$

$$d_{c,t}^C = a_c^C m_{c,t}^J \left( (\Gamma_c)^{\alpha_c^k} - 1 \right), \ \forall c \in \mathbb{C} \tag{1.h}$$

$$d_u = F_u(e_u), \forall u \in \mathbb{U} \tag{1.i}$$

$$0 \le e_u, \forall u \in \mathbb{U} \tag{1.j}$$

$$\sum_{g \in \mathbb{G}} v_g^G + \sum_{p \in \mathbb{P}_j^{to}} m_{p,n_p^{seg}}^t - \sum_{p \in \mathbb{P}_j^{fr}} m_{p,1}^t = \sum_{l \in \mathbb{L}_j} d_l$$

$$+ \sum_{u \in U_j} d_u + \sum_{c \in \mathbb{C}_j} d_{c,t}^C, \forall j \in \mathbb{J}, \ t \in \mathbb{T} \tag{1.k}$$

$$\sum_{p \in \mathbb{P}} \sum_{s \in \mathbb{S}_p, t = n_T} (\rho_{p,s}^t \pi D_p^2 \Delta x / 8\Delta t) \le L_{\min} \tag{1.l}$$

The natural gas system dynamics are described by one-dimensional nonlinear hyperbolic partial differential equations, and the partial differential equations are further transformed into algebraic equations in (1.a)-(1.c) by the finite difference method, in which the time step $\Delta t$ is set to 5 minutes. Constraint (1.c) includes the initial gas density $\rho_{p,s}^0$, which reflects the limitations of the initial state of natural gas systems. The gas density at each node is limited by (1.d). The natural gas density relationship at each node is described in (1.e) and (1.f). The compressor model is presented in (1.g) and (1.h). Constraint (1.g) means that the density is increased up to $\Gamma_j$ times in the compressor nodes, in which $\Gamma_j$ is a known parameter in intraday operation. Constraint (1.h) describes the relationship between the gas consumption of the compressor and the gas flow through the compressor. Constraints (1.i) and (1.j) describe the relationship between the output and gas withdrawal of gas-fired units, in which $F_u(\bullet)$ is a linear function. The natural gas supply-demand balance for each node is ensured by (1.k), in which the gas wells and traditional gas loads are modeled with a fixed dispatch scheme. Constraint (1.l) limits the minimum total linepack level of the natural gas system.

The DSR $\Omega_{DSR}$ can be portrayed as a closed region in which the operation point satisfies the dynamic natural gas system constraints (1.a)-(1.l). As mentioned above, the demand of traditional gas loads and output of gas wells is known and fixed during intraday operation, so only the gas withdrawal of gas-fired units needs to be evaluated in the DSR, and $\Omega_{DSR}$ is formulated as:

$$\Omega_{DSR} = \left\{ \boldsymbol{x} / f(\boldsymbol{x}, \boldsymbol{y}) = \boldsymbol{0}, g(\boldsymbol{x}, \boldsymbol{y}) \le \boldsymbol{0} \right\} \tag{1.m}$$

where $\boldsymbol{x}$ is the vector of the gas-fired unit gas withdrawals and $\boldsymbol{y}$ is a vector of the remaining variables except $\boldsymbol{x}$. $f(\cdot)$ denotes the equality constraints (1.a)-(1.c), (1.e)-(1.i), (1.k). $g(\cdot)$ denotes the inequality constraints (1.d), (1.j), and (1.l).

It occurs naturally to us that the DSR can be portrayed by using the boundary of the gas-fired unit gas withdrawal. Then, the evaluation for the DSR $\Omega_{DSR}$ with multiple natural gas-fired units can be converted to the following vector optimization problem, where the boundary of the DSR is constructed by the Pareto front.

*vector optimization problem for the upper boundary*

$$\min \ -\boldsymbol{x} \tag{1.n}$$

$$Subject \ to \ (1.a)\text{-}(1.l)$$

*vector optimization problem for the lower boundary*

$$\min \ \boldsymbol{x} \tag{1.o}$$

$$Subject \ to \ (1.a)\text{-}(1.l)$$

To visualize the relationship between the DSR and gas-fired unit gas withdrawal boundary, the 2-dimensional DSR for the situation where two gas-fired units are located at nodes $i$ and $j$, is portrayed in Fig. 3.

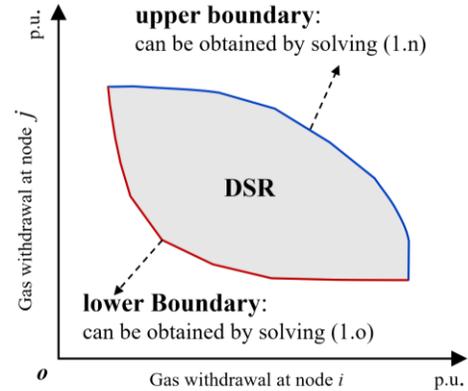

Fig. 3 Two-dimensional DSR including two gas-fired unit gas withdrawal nodes.

In practice, the gas-fired units may be located at multiple nodes of natural gas systems, which means that the above vector optimization problem for the DSR (1.n) and (1.o) is a high-dimensional optimization problem, and it is difficult to obtain the Pareto front (i.e., the boundary of the DSR). Although it has been proposed in [28] that the vector optimization problem can be transformed into a single-objective optimization by enumerating the weight coefficients of the elements in $\boldsymbol{x}$, the traversal of the weight coefficients is also challenging. Therefore, this paper proposes a dimensionality reduction evaluation method for DSR based on the adjusted regulation of the gas-fired units and the power-gas coupling relationship in IEGSs.

### C. Dimensionality reduction evaluation for DSR

As mentioned in **Assumption-2**, there is a linear relationship between the adjustment amount of the gas-fired unit gas withdrawal, which is shown in (2.a) and (2.b). Although the linear relationship limitation increases the number of constraints, as we analyze below, it is also able to reduce the high-dimensional problem to a one-dimensional search problem, which is a significant change.

$$d_u = \hat{d}_u^G + \Delta d_u^G, \forall u \in \mathbb{U} \tag{2.a}$$

$$\Delta d_u^G = \beta_u d^G, \forall u \in \mathbb{U} \tag{2.b}$$

The intraday gas-fired unit gas withdrawals and the adjustment amounts of gas withdrawal are described in (2.a) and (2.b), respectively. $\Delta d_u^G$, $\hat{d}_u^G$ and $d_u$ are the adjustment amount, dispatch scheme, and actual amount of gas withdrawal for gas-fired unit $u$, respectively. $d^G$ is the total adjustment amount of gas withdrawal. The participation factor $\beta_u$ is known from power systems.

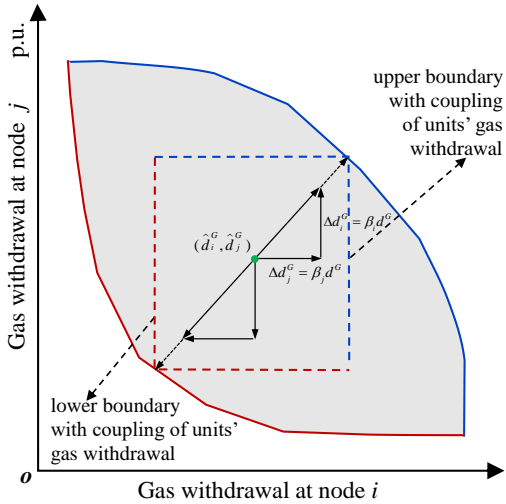

Fig. 4 Searching for the DSR boundary with two coupled gas-fired units' gas withdrawals in a 2-dimensional DSR.

The gas-fired unit gas withdrawals are adjusted according to the given participation factors $\beta_u$ and $d^G$. As shown in Fig. 4, the direction of the gas withdrawal change is fixed by $\beta_j$ and $\beta_i$, and the magnitude of the gas withdrawal change is determined by $d^G$. Consequently, for any IEGS with a given participation factor $\beta_u$, the upper and lower boundaries of the DSR can be constructed using $d^G$. It should be noted that to visualize the effect of the gas withdrawal coupling for gas-fired units, a two-dimensional DSR is used in Fig. 4, and it is straightforward to see that the above conclusions also hold for high-dimensional DSRs. In light of the above considerations, the evaluation functions (2.c) and (2.d) for DSRs are constructed. Then, the high-dimensional optimization problem is converted into a one-dimensional search problem for $d^G$. The upper and lower boundary models of the DSR in this paper are given below.

**DSR Model:**

*upper boundary model of DSR (UB Model)*

$$\min -d^G \tag{2.c}$$

*Subject to* (1.a)-(1.l) and (2.a)-(2.b)

*lower boundary model of DSR (LB Model)*

$$\min d^G \tag{2.d}$$

*Subject to* (1.a)-(1.l) and (2.a)-(2.b)

***Validity analysis of the dimensionality reduction method:*** By solving the UB Model and LB Model, the maximum upward and downward total adjustment amount for the gas-fired unit gas withdrawals can be found. Then, the maximum total adjustment $\Delta d_u^G$ is allocated to each gas-fired unit according to (2.b), so the boundary of gas withdrawal for every gas-fired unit is obtained according to (2.a). Finally, as shown in Fig. 5, the DSR that appears in the 2-dimensional space is constructed using the upper boundary and lower boundary of every gas-fired unit's gas withdrawal. All the gas withdrawal points inside the region are secure.

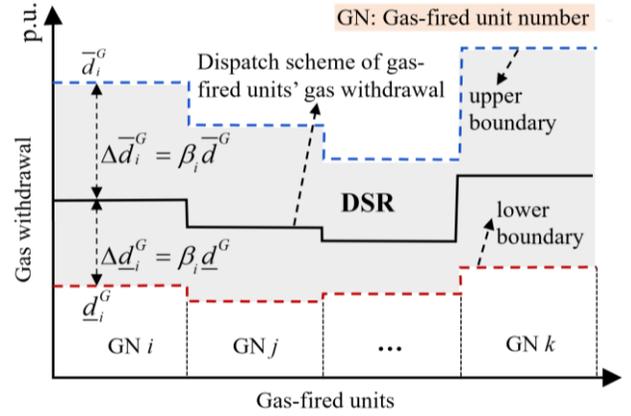

Fig. 5 Visualization of DSR with multiple gas-fired units' gas withdrawal nodes using our dimensionality reduction method. The gray area is the DSR of the natural gas system, $\overline{d}^G$, $\underline{d}^G$ are the maximum upward and downward total adjustment for the gas-fired unit gas withdrawals, respectively, and $\overline{d}_i^G$, $\underline{d}_i^G$ are the upper boundary and lower boundary of the gas-fired unit's withdrawal at node $i$.

## 4. Fast evaluation algorithms for DSR

Most constraints of DSR models (UB Model and LB Model) are linear and easy to handle, and nonconvexity only exists in constraint (1.a), which, however, makes DSR models nonconvex, NP-hard, and difficult to solve directly. Semidefinite relaxations can transform nonconvex natural gas system models into convex models [29], which provides a way to solve DSR models. However, the tightness of the semidefinite program (SDP) cannot be guaranteed, so the solution of the SDP may be physically infeasible. Recently, a rank minimization algorithm (RMA) was proposed in [13], which significantly improved the tightness of the SDP. Inspired by RMA, this paper proposes the FE to solve DSR models.

The idea of the FE is shown in Fig. 6. First, the feasibility convexification algorithm based on RMA is utilized to obtain rank-one solutions for the SDP. Then, an optimal bisection algorithm is proposed to address problems for which the RMA fails to yield a rank-one optimal solution. The processes of the RMA for the UB Model and LB Model are similar, so only the process for the UB Model is described, and the process for the LB Model is not repeated in this paper. The detailed FE is formulated as follows.

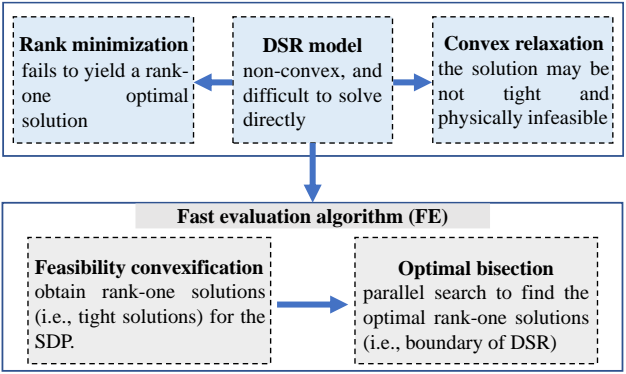

Fig. 6 Idea and structure of the FE for DSR models

### A. Feasibility convexification algorithm

First, the auxiliary variables $\gamma_{p,s}^{t+1}$ are introduced to relax the nonconvex constraint (1.a) into linear constraint (3.a) and SDP constraint (3.b).

$$\frac{\rho_{p,s+1}^{t} - \rho_{p,s}^{t}}{\Delta x} + \frac{\gamma_{p,s}^{t}}{\pi^2 D^5/8f} = 0,$$
$$\forall p \in \mathbb{P}, \ s \in \mathbb{S}_p, t \in \mathbb{T} \tag{3.a}$$

$$\begin{bmatrix} \gamma_{p,s}^{t} & m_{p,s}^{t} \\ m_{p,s}^{t} & \rho_{p,s}^{t} \end{bmatrix} \succeq 0, \forall p \in \mathbb{P}, \ s \in \mathbb{S}_p, t \in \mathbb{T} \tag{3.b}$$

As previously mentioned, the SDP relaxation (3.b) may not be tight. By observing (3.b), a natural conjecture that arises is that the smaller $\gamma_{p,s}^{t}$ is, the closer the determinant of the SDP matrix may be to zero, which means that it is more likely to obtain a rank-one solution, i.e., the relaxation is tight. Although there is no rigorous theoretical proof that minimizing $\gamma_{p,s}^{t+1}$ can ensure the tightness of the solution, the verified results of [13] have shown that this method usually performs well.

Based on [13], the objective function (3.c) containing only $\gamma_{p,s}^{t+1}$ is constructed to minimize the rank of the SDP constraints that ensure that the SDP is tight, i.e., the solution is physically feasible. Then, the UB Model is transformed into the convex DSR-I Model.

### DSR-I Model:

$$\min \sum_{t \in \mathbb{T}} \sum_{p \in \mathbb{P}} \sum_{s \in \mathbb{S}_p} \gamma_{p,s}^{t} \tag{3.c}$$

*Subject to,* (1.a)-(1.l), (2.a)-(2.b), (3.a)-(3.b)

### B. Optimal bisection algorithm

It should be noted that to guarantee the tightness of SDP, the objective function (3.c) containing only $\gamma_{p,s}^{t}$ replaces the original objective function (2.c), which makes the DSR-I Model fail to find the optimal solution for the UB Model (i.e., the upper boundary of the DSR). For this reason, an optimal bisection algorithm in Table 1 that can find high-quality feasible solutions is proposed.

First, the optimal constraint (4.a) is added to the DSR-II Model to limit $d^G$ by $\eta$. The smaller $\eta$ is, the smaller the upper limit for $-d^G$, and the closer the solution of the DSR-II Model is to the optimal solution of the UB Model. Therefore, by modifying the value of the given parameter $\eta$, the optimal solution of the UB Model can be found by the DSR-II Model.

### DSR-II Model:

Min (3.c)

*Subject to* (3.a)-(3.b), (1.b)-(1.l), (2.a)-(2.b)

$$-d^G \leq \eta \tag{4.a}$$

Second, a bisection algorithm with parallel processes is proposed to quickly search for the optimal $\eta$ (corresponding to the minimum $-d^G$ of the UB Model and the upper boundary of the DSR). The precise description of the optimal bisection algorithm is shown in Table 1.

**Table 1** Optimal bisection algorithm for DSR

| **Optimal Bisection Algorithm** |
|---|
| **1** Set the stopping tolerance $\varepsilon$ and number of parallel processes $N$. Set $k=0$ |
| **2** Solve the relaxed UB Model with constraints (3.a) and (3.b) to obtain $\eta_{\min}$. <br> Let $\eta = \eta_{\min}$, and solve the DSR-II Model. |
| **3** **if** the DSR-II Model's solution is rank-one. <br>      **Output** the solution $x_{k,i}$. <br>      **else**    Let $\eta_{\min,k} = \eta_{\min}, \ \eta_{\max,k} = 0$. |
| **4** **Repeat** <br>      Set parallel sampling points for $\eta$: Select $N$ sampling points <br>      $\eta_{k,i} = \eta_{\min,k} + (\eta_{\max,k} - \eta_{\min,k}) * i / (N-1), i = 1,2,\cdots,N$. <br>      Parallel solving the DSR-II model with $\eta_{k,i}$, and denote by <br>      $x_{k,i}$ the associated solution of the DSR-II model, in which <br>      $i = 2,3,\cdots,N-1$. |
| **5** **Update**   $\eta_{\max,k+1} = \eta_{k,\hat{i}}$, in which $\hat{i} = \min\{i \mid \operatorname{rank}(x_{k,i}) \leq 1, i = 1,2,\cdots,N\}$ <br>                $\eta_{\min,k+1} = \eta_{k,\hat{i}-1}$ <br>                $k = k+1$; |
| **6** **Until**   $\eta_{\max,k} - \eta_{\min,k} \leq \varepsilon$ |
| **7** **Output**   $x_{k-1,\hat{i}}$ |

$\operatorname{rank}(x_{k,i})$ denotes the rank of the SDP constraints at the solution $x_{k,i}$.

The search range selection for $\eta$ and the termination criteria for the optimal bisection algorithm are given as follows.

**Range selection:** $\eta$ is a parameter that denotes a possible value of the function (2.c). Denote by $\eta*$ the optimal value of the UB Model, which corresponds to the upper boundary of the DSR. It is obvious that $\eta* \leq 0$. Meanwhile, replace the nonconvex constraint (1.a) in the UB Model with the relaxed constraints (3.a) and (3.b) and solve the relaxed model directly. Denote by $\eta_{\min}$ the associated value of the function (2.c) in this case; it is easy to see that $\eta^* \geq \eta_{\min}$. Obviously, the initial range of $\eta$ is $[\eta_{\min}, 0]$, and $\eta*$ can be found by searching for the minimum $\eta$ within this range with feasible solutions of the DSR-II model.

**Termination criteria:** In the event that the bisection algorithm fails to yield a rank-one solution of the DSR-II Model, one of two reasons may be at play. First, the given $\eta$ may be so small that no rank-one solutions exist for the DSR-II Model. Second, there may be a rank-one solution of the DSR-II Model, but the RMA fails to recover it, as the RMA is an approximate algorithm and is not always guaranteed to recover the rank-one solution. Considering these two scenarios, **the minimum $\eta$ for the DSR-II Model with a rank-one solution is regarded as** $\eta*$ (which corresponds to $d^G$ as the total adjustment amount of gas withdrawal for the upper boundary of the DSR). The specific process is shown in steps 5 and 6 in Table 1. In this way, the optimality and feasibility of the bisection algorithm are balanced.

## 5. Case studies

In this section, 10-node and 67-node natural gas systems are constructed respectively to verify the necessity and effectiveness of the DSR model and FE proposed in this paper,. The configuration of the 10-node test system is displayed in Fig. 7. The gas-fired units are located at nodes 8 and 10, and the other parameters of the test system can be found in Table B1 in Appendix B. The 67-node natural gas system is constructed based on the test system in [31]. The gas-fired units are distributed over 13 nodes in the 67-node natural gas system, the specific locations of the gas-fired units and the corresponding participation factors are shown in Table B2 in Appendix B, and the rest of the detailed parameters can be found in [31]. Considering the time granularity of the intraday dispatch scheme, the time interval of evaluation for DSR is 15 minutes, with a time step of 5 minutes. All the code is implemented on a PC with a 2.90 GHz Intel(R) Core(TM) i7 CPU and 64-bit Windows operating system. The original nonconvex model is solved directly using IPOPT, while the DSR-II model used in the proposed FE is solved by MOSEK [30].

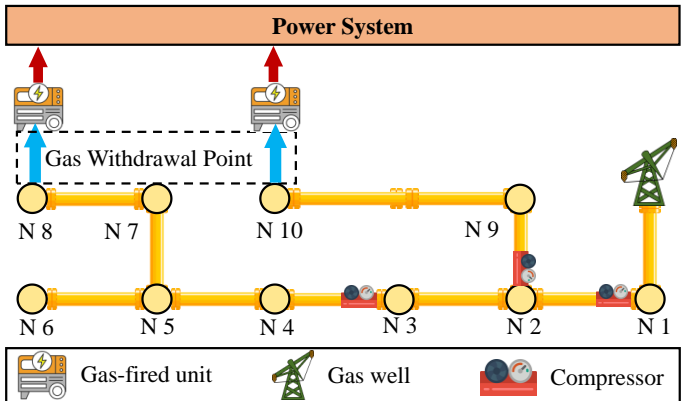

Fig. 7 Schematic diagram of the 10-node natural gas system

### A. Necessity verification for DSR

To verify the necessity of the DSR model, the DSR and SSR are evaluated for the 10-node system at different times. The SSR is constructed according to the method in [15], with the addition of constraints (2.a) and (2.b) to account for the coupling relationship between the gas withdrawals of gas-fired units. Specifically, the DSR and SSR are evaluated at three time points, T1, T2, and T3, where the gas loads are 112 kg/s, 126 kg/s, and 140 kg/s, respectively. The DSR and SSR at T1, T2, and T3 are shown in Fig. 8.

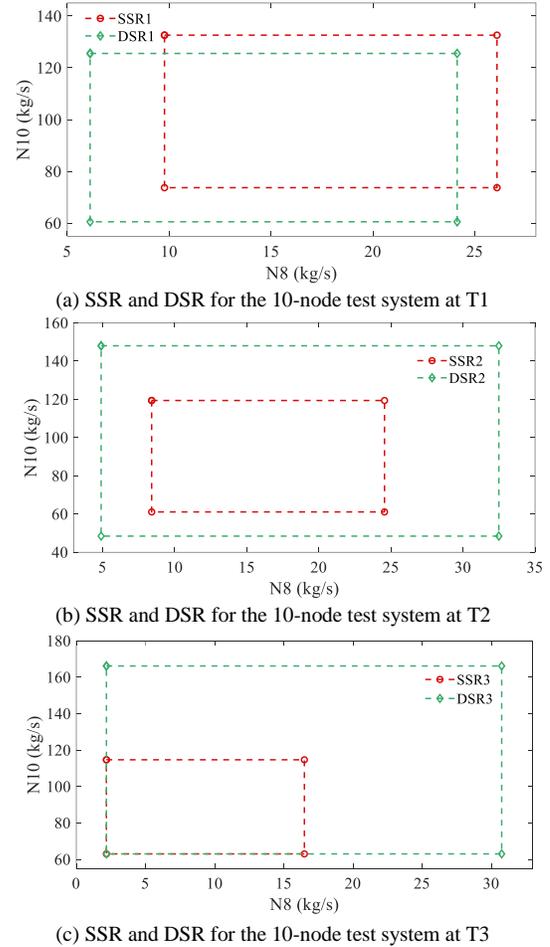

(a) SSR and DSR for the 10-node test system at T1

(b) SSR and DSR for the 10-node test system at T2

(c) SSR and DSR for the 10-node test system at T3

Fig. 8 SSR and DSR for the 10-node natural gas system at different times

As mentioned in **Assumption-2**, it is known that there is a linear relationship based on AGC participation factors between the adjustment amounts of the gas-fired unit gas withdrawals. As shown in Fig. 8, considering the gas withdrawal characteristics of the gas-fired units mentioned above, the DSR corresponding to the secure gas withdrawal range of the gas-fired units in the 10-node system with only two power-gas coupling nodes is a rectangle.

As indicated by Fig. 8, there can be significant differences in the ranges of the DSR and SSR, e.g., the two can intersect (see Fig. 8a), the SSR can be completely located within the DSR (see Fig. 8b), or the SSR can share a common boundary with DSR (see Fig. 8c). This means that the SSR is not always a conservative estimate of the DSR (see Fig. 8a) because practical dynamic operational constraints must be taken into account by the DSR. Additionally, the DSR also exhibits certain differences at different times.

Moreover, if the SSR or an outdated DSR is adopted in power system scheduling, the dispatch instructions issued based on the SSR or outdated DSR at some point in time may cause the gas-fired units to extract gas outside the security operation region, resulting in security issues in the natural gas system. The following example based on time T1 illustrates the problem.

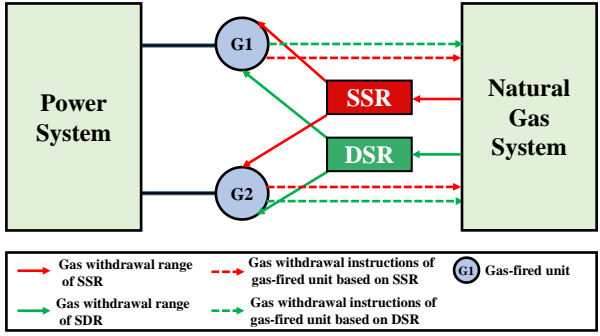

Fig. 9 Schematic diagram of DSR- and SSR-based electricity-gas system coordinated dispatch.

As shown in Fig. 9, the DSR and SSR are sent to the power system, respectively. Subsequently, the power system reformulates the dispatch scheme based on the DSR and SSR, and sends scheduling instructions for gas withdrawal of gas-fired units to the natural gas system. The operating state of the natural gas system is simulated according to the gas withdrawal scheme of the gas-fired units based on the DSR and SSR. The dispatch scheme and simulation results are shown in Fig. 10 and Table 2.

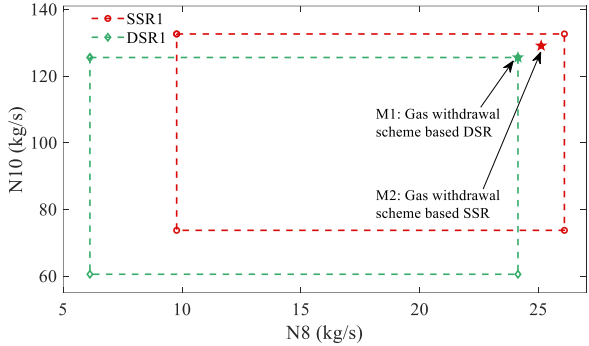

Fig. 10 Scheduling instructions of gas withdrawal of gas-fired units

**Table 2** Security validation of the gas withdrawal point

| Gas withdrawal point | M1 | M2 |
|---|---|---|
| **Gas withdrawal** (kg/s) | (24.13, 125.56) | (25.1, 129.09) |
| **Security** | secure | insecure |

Fig. 10 illustrates that the secure operation region of the natural gas system at time T1 is overestimated by the SSR. Consequently, the scheduling instructions M2 based on the SSR exceed the secure gas supply range of the natural gas system, resulting in an insecure state of the natural gas system, as shown in Table 2. The DSR model proposed in this paper accurately describes the secure operation ranges of the natural gas system with dynamics, thus ensuring the security of the system. This highlights the security benefits of utilizing the DSR while underlining the security risks associated with the SSR. Similarly, adopting an outdated DSR can also lead to security issues, emphasizing the necessity of dynamic updates for DSRs.

### B. Visualization of a high-dimensional DSR

As demonstrated by the following 67-node test system, there may be multiple power-gas coupling nodes in the IEGS, and it can be extremely difficult to portray the high-dimensional DSR in such cases. In this paper, the high-dimensional DSR is mapped to each gas-fired unit through the proposed AGC-based high-dimensional mapping projection method to obtain a visualized DSR. As shown in Fig. 11, based on the coupling equations (2.a) and (2.b), the total adjustment amount of gas withdrawal is allocated to each gas-fired unit.

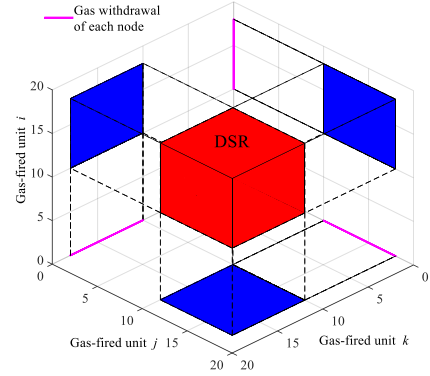

Fig. 11 Schematic diagram of the mapping projection method for a high-dimensional DSR.

The 67-node test system with 13 power-gas coupling nodes is used to illustrate the effectiveness of the visualization method. The visualized representation of the high-dimensional DSR of the 67-node test system is shown in Fig. 12. The gas withdrawal range obtained by the high-dimensional mapping projection method forms a two-dimensional DSR that is more suitable for collaborative operation in IEGS than the high-dimensional SR.

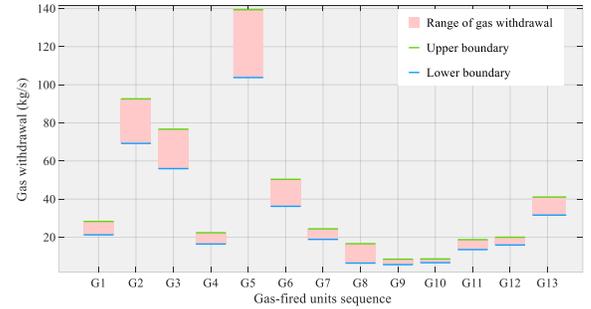

Fig. 12 Gas withdrawal range for each gas-fired unit in the 67-node test system.

### C. Performance analysis for FE

#### 1) Validity verification for FE.

To verify the necessity of the proposed FE, the solution of the UB Model is taken as an example, and IPOPT and FE are compared by adopting them to solve the DSR model in the 10-node system and 67-node system, respectively. The results of IPOPT and FE are shown in Table 3.

**Table 3** Comparison of results of the 10-node and 67-node natural gas system for IPOPT and FE

| | **Method** | Solution time (s) | Relaxation ratio |
|---|---|---|---|
| **10-node** | **Solved by IPOPT** | 0.15 | -- |
| | **FE** | 0.45 | 8.0 |
| **67-node** | **Solved by IPOPT** | ≥300 | -- |
| | **FE** | 3.45 | 6.3 |

IPOPT solves the original nonconvex model directly without a relaxation gap, so the solution obtained by IPOPT does not have a relaxation ratio.

The relaxation ratio in the last column of Table 3 is the logarithmic ratio of the largest and second-largest eigenvalues of the SDP matrix, and a larger relaxation ratio corresponds to tighter relaxation [13]. The relaxation ratio of FE exceeds 6 in all cases, which can be considered evidence that FE achieves a tight relaxation of the nonconvex model based on the criteria in [13].

It can be seen from Table 3 that in the 10-node test system, both IPOPT and FE solve the DSR model quickly. However, as the scale of the natural gas system expands, the solution time of IPOPT increases significantly. IPOPT fails to completely solve the problem within 300 seconds in the 67-node test system. As mentioned above, during intraday operation, the granularity of the dispatch scheme of the power system is usually 5 or 15 minutes. It is evident that IPOPT cannot meet the dynamic updating requirements of DSRs. In contrast, the FE proposed in this paper exhibits better performance by completing the DSR evaluation process within a mere 3.45 seconds (over 80 times faster than IPOPT), highlighting its potential to significantly enhance the evaluation speed of DSRs, which is crucial to meeting the dynamic update requirements of DSRs during IEGS intraday operation.

*2) Solution stability verification for FE.*

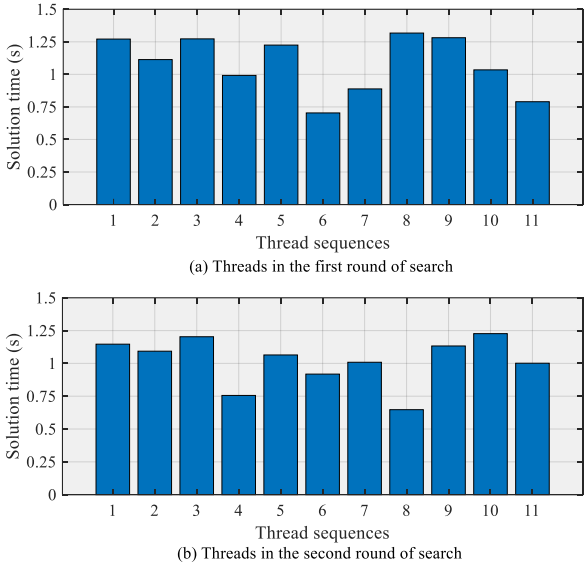

Fig. 13 The solution time of every thread of the FE

As demonstrated in Table 1, a parallel computing scheme is adopted in the FE, and eleven parallel threads simultaneously solve the DSR-II Model with different values of $\eta$ in one round of search. To illustrate the stability and effectiveness of FE, the solution time of each parallel thread of the 67-node system is shown in Fig. 13. The boundary of the DSR is found after 2 rounds of parallel search. Moreover, the maximum solution time among all threads is only 1.28 seconds, indicating that the convex SDP-based FE is stable and efficient.

*D. Sensitivity analysis for DSR obtained by AGC participation factors*

As shown in Section 2.C, the range of the DSR is directly related to the selection of AGC participation factors of the gas-fired units. In this part, the 10-node test system with three different participation factors is set to investigate the effect of the participation factors on the DSR. Specifically, first, the participation factors in Section 5.A are directly adopted, and the DSR obtained is denoted by DSR$^{AGC1}$. Then, the participation factors of the gas-fired units at node 10 is increased and decreased by 20%, and the corresponding DSRs are denoted by DSR$^{AGC2}$ and DSR$^{AGC3}$. The specific participation factors are given in Appendix B.

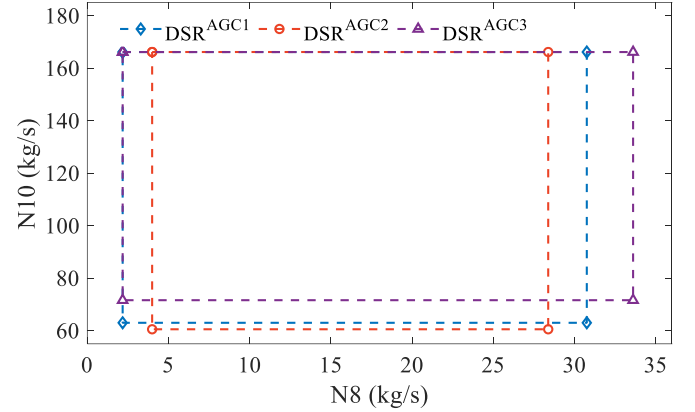

Fig. 14 DSRs for the 10-node test system with different participation factors.

As expected, there are significant differences between the DSRs with different AGC participation factors. Fig. 14 shows that the upper boundary of the DSR gradually increases with the relative decrease in the participation factors of the gas-fired units at node 10, which means that the gas-fired units obtain more gas supply. The reason for this is that the gas-fired units mainly withdraw gas from node 10, where the gas withdrawal usually approaches the upper limit more quickly than at node 8. Therefore, the upper limit of gas withdrawal at node 10 becomes an important factor limiting the upper boundary of the DSR. By appropriately decreasing the participation factors of the gas-fired units at node 10 (equivalent to increasing the proportion of participation factors of gas-fired units at node 8), the gas withdrawal at node 8 corresponding to the upper boundary of the DSR increases when the gas withdrawal at node 10 reaches its upper limit. The change in the lower boundary of the DSR is similar to that of the upper boundary.

The total gas withdrawal areas of DSR$^{AGC1}$, DSR$^{AGC1}$, and DSR$^{AGC1}$ are shown in Table 4. As the participation factors change, the total gas withdrawal (i.e., regulation capacity) of the gas-fired units located at different nodes also changes. This implies that the participation factors can be further optimized to achieve the largest total gas withdrawal area, and this content will be left to future work.

**Table 4** Comparison of areas of different DSRs

| DSRs | DSR$^{AGC1}$ | DSR$^{AGC2}$ | DSR$^{AGC3}$ |
|---|---|---|---|
| **Area** (kg$^2$/s$^2$) | 294.8 | 257.6 | 297.1 |

## 6. Conclusion

This paper introduces the concept of the DSR, which accurately reflects the dynamic characteristics of the natural system in contrast to the traditional SSR and can effectively ensure the secure operation of natural gas systems and IEGSs.

Specifically, to address some of the key challenges in DSR evaluation, this paper makes the following contributions: (1) the concept and model of DSR, which consider dynamic characteristics of the natural gas system, are proposed; (2) a dimensionality reduction method is proposed to efficiently solve and visualize the high-dimensional DSR evaluation model; and (3) a fast evaluation algorithm is developed to address the difficulties resulting from nonconvex dynamic constraints in the DSR model.

The simulation results demonstrate that DSRs can effectively ensure the secure operation of natural gas systems and IEGSs while highlighting the potential security risks posed by SSRs. The results from a relatively large system with multiple power-gas coupling nodes show that the dimensionality reduction method helps visualize the DSR by mapping the projection of the gas withdrawal in a high-dimensional DSR to each gas-fired unit. The results from the relatively large system also indicate that the FE proposed in this paper significantly reduces computation time by a factor of 80 compared to that of the nonconvex method with IPOPT. Moreover, the results from different AGC participation factors clarify their important impact on the range of the DSR.

In the future, further research will be conducted on the application of DSR to the coordinated operation and preventative control of the IEGS. Furthermore, in addition to the rank minimization approach in this paper, other rank minimization approaches, such as those in [32] and [33], will also be explored for potential enhancements in future work.

## Acknowledgment

This work was supported by the National Natural Science Foundation of China under Grant 52007105.

## Appendix A

**Table A1** Parameters of the pipeline in the 3-node natural gas system

| Pipeline number | From node | To node | Length (km) | Diameter (m) |
|---|---|---|---|---|
| 1 | 3 | 1 | 60 | 0.868 |
| 2 | 1 | 2 | 80 | 0.868 |

**Table A2** Parameters of the gas well in the 3-node natural gas system

| Number | Node | Minimum output (kg/s) | Maximum output (kg/s) |
|---|---|---|---|
| S1 | 3 | 120 | 220 |

## Appendix B

**Table B1** Parameters of the pipeline in the 10-node natural gas system

| Pipeline number | From node | To node | Length (km) | Diameter (m) |
|---|---|---|---|---|
| 1 | 1 | 2 | 100 | 0.868 |
| 2 | 2 | 3 | 150 | 0.868 |
| 3 | 3 | 4 | 50 | 0.868 |
| 4 | 4 | 5 | 15 | 0.603 |
| 5 | 5 | 6 | 10 | 0.603 |
| 6 | 5 | 7 | 5 | 0.603 |
| 7 | 7 | 8 | 10 | 0.603 |
| 8 | 2 | 9 | 5 | 0.868 |
| 9 | 9 | 10 | 60 | 0.868 |

**Table B2** Locations and participation factors of gas-fired units in the 10-node test system

| Gas-fired unit number | Node number | Participation factor of gas-fired units | | |
|---|---|---|---|---|
| | | AGC1 | AGC2 | AGC3 |
| G1 | 10 | 0.0782 | 0.0811 | 0.0750 |
| G2 | 10 | 0.0735 | 0.0763 | 0.0705 |
| G3 | 10 | 0.0758 | 0.0786 | 0.0726 |
| G4 | 10 | 0.1439 | 0.1493 | 0.1379 |
| G5 | 10 | 0.1397 | 0.1449 | 0.1339 |
| G6 | 10 | 0.1421 | 0.1475 | 0.1362 |
| G7 | 10 | 0.0054 | 0.0056 | 0.0051 |
| G8 | 10 | 0.0122 | 0.0127 | 0.0117 |
| G9 | 10 | 0.0039 | 0.0040 | 0.0038 |
| G10 | 10 | 0.0042 | 0.0044 | 0.0040 |
| G11 | 10 | 0.0053 | 0.0055 | 0.0051 |
| G12 | 10 | 0.0988 | 0.1025 | 0.0947 |
| G13 | 8 | 0.0417 | 0.0360 | 0.0479 |
| G14 | 8 | 0.0371 | 0.0321 | 0.0427 |
| G15 | 8 | 0.0327 | 0.0283 | 0.0376 |
| G16 | 8 | 0.0414 | 0.0358 | 0.0476 |
| G17 | 8 | 0.0288 | 0.0248 | 0.0331 |
| G18 | 8 | 0.0354 | 0.0306 | 0.0407 |

**Table B3** Locations and participation factors of gas-fired units in the 67-node test system

| Gas-fired unit number | Node number | Participation factor of gas-fired units |
|---|---|---|
| G1 | 4 | 0.0482 |
| G2 | 6 | 0.1600 |
| G3 | 9 | 0.1417 |
| G4 | 13 | 0.0406 |
| G5 | 17 | 0.2438 |
| G6 | 23 | 0.0969 |
| G7 | 31 | 0.0377 |
| G8 | 44 | 0.0696 |
| G9 | 49 | 0.0195 |
| G10 | 50 | 0.0135 |
| G11 | 55 | 0.0357 |
| G12 | 59 | 0.0280 |
| G13 | 64 | 0.0648 |